\documentclass[journal,twocolumn]{IEEEtran}

\usepackage{times}
\usepackage{epsfig}

\usepackage{cite}
\usepackage{amssymb}
\usepackage{amsmath}
\usepackage{graphicx}
\usepackage{graphics}
\usepackage{textcomp}
\usepackage{subfigure}
\usepackage{xspace}
\usepackage{arydshln}
\usepackage{sidecap}
\usepackage{caption}
\usepackage[]{algorithm2e}

\usepackage{enumitem}
\usepackage[bookmarks=false]{hyperref}
\usepackage{cleveref}
\usepackage{multirow}
\usepackage{tabularx}
\usepackage{epstopdf}
\usepackage{float}

\usepackage{color,soul}
\usepackage[dvipsnames]{xcolor}
\soulregister\cite7
\soulregister\ref7

\DeclareGraphicsExtensions{.eps,.pdf,.jpeg,.png}

\makeatletter

\newcommand{\Rmnum}[1]{\expandafter\@slowromancap\romannumeral #1@}

\newtheorem{thm}{Theorem} 
\newtheorem{defn}[thm]{Definition} 
\makeatother

\def\x{{\mathbf x}}

\def\z{{\mathbf z}}

\begin{document}

\title{Better Compression with Deep Pre-Editing}

\author{Hossein Talebi, Damien Kelly, Xiyang Luo, Ignacio Garcia Dorado, \\ Feng Yang, Peyman Milanfar and Michael Elad \\ Google Research - Mountain View, California\\
}

\maketitle

\graphicspath{{Figures/}}

\begin{abstract}

Could we compress images via standard codecs while avoiding visible artifacts? The answer is obvious -- this is doable as long as the bit budget is generous enough. What if the allocated bit-rate for compression is insufficient? Then unfortunately, artifacts are a fact of life. Many attempts were made over the years to fight this phenomenon, with various degrees of success. In this work we aim to break the unholy connection between bit-rate and image quality, and propose a way to circumvent compression artifacts by pre-editing the incoming image and modifying its content to fit the given bits. We design this editing operation as a learned convolutional neural network, and formulate an optimization problem for its training. Our loss takes into account a proximity between the original image and the edited one, a bit-budget penalty over the proposed image, and a no-reference image quality measure for forcing the outcome to be visually pleasing. The proposed approach is demonstrated on the popular JPEG compression, showing savings in bits and/or improvements in visual quality, obtained with intricate editing effects.  
\end{abstract}


\section{Introduction}
\label{sec:introduction}

Commonly used still image compression algorithms, such as JPEG \cite{wallace1992jpeg}, JPEG-2000 \cite{christopoulos2000jpeg2000}, HEIF \cite{heif2013} and WebP \cite{webp2012} produce undesired artifacts when the allocated bit rate is relatively low. Blockiness, ringing, and other forms of distortion are often seen in compressed-decompressed images, even at intermediate bit-rates. As such, the output images from such a compression procedure are of poor quality, which may hinder their use in some applications, or more commonly, simply introduce annoying visual flaws. 

Numerous methods have been developed over the years to confront this problem. In Section \ref{sec:Prior} we provide a brief review of the relevant literature, encompassing the various strategies taken to fight compression artifacts. Most of the existing solutions consider a post-processing stage that removes such artifacts after decompression  \cite{zakhor1992iterative,shen1998review,chen2001adaptive,triantaffilidis2002blockiness,alter2005adapted,averbuch2005deblocking,kartalov2007adaptive,weiss2008compression,zhai2009efficient,jung2012image,zhang2013compression,dar2016postprocessing,du2007post,du2011new,kwon2015efficient}. Indeed, hundreds of papers that take this post-processing approach have been published over the years, including recent deep-learning based solutions (e.g., \cite{Dong2015deep,Wang2016deep,Galteri2017deep,Cavigelli2017deep}).  

\begin{figure}[!t]
\vspace{-0 mm}
\begin{center}
\subfigure[\scriptsize Input]{
    \includegraphics*[viewport=170 480 280 610, scale=1.0]{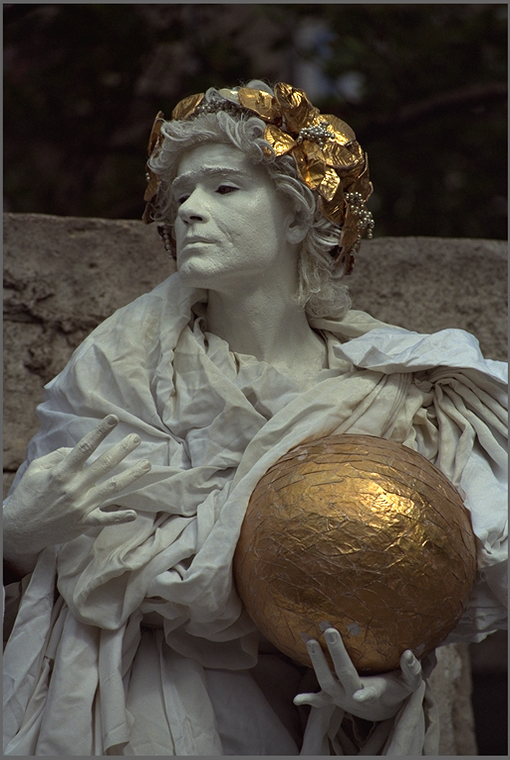}
}
\subfigure[\scriptsize Baseline JPEG (0.4809 bpp)]{
    \includegraphics*[viewport=170 480 280 610, scale=1.0]{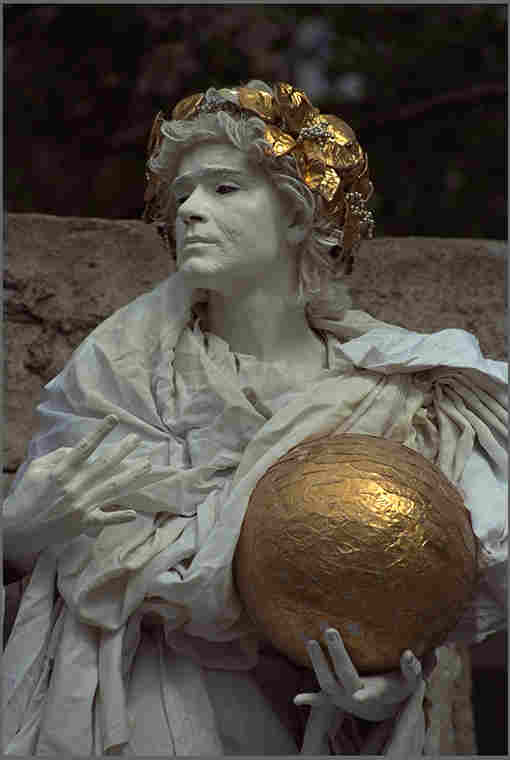}
}
\subfigure[\scriptsize Edited input]{
\includegraphics*[viewport=170 480 280 610, scale=1.0]{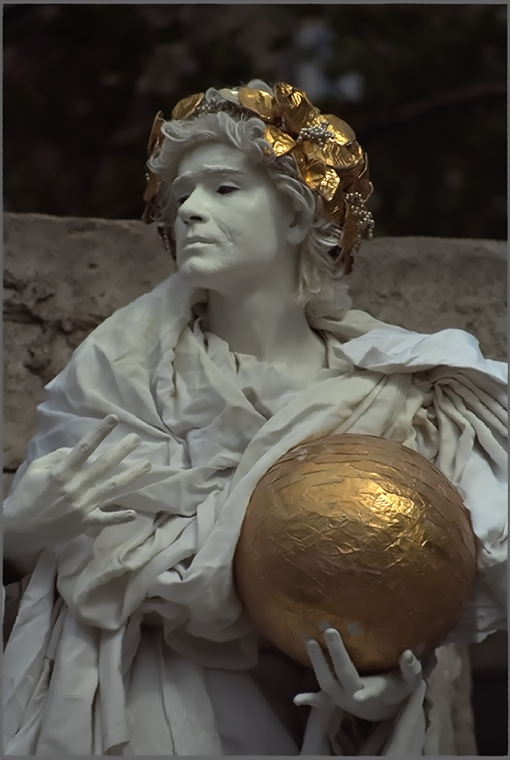}
}
\subfigure[\scriptsize JPEG after editing (0.4726 bpp)]{
\includegraphics*[viewport=170 480 280 610, scale=1.0]{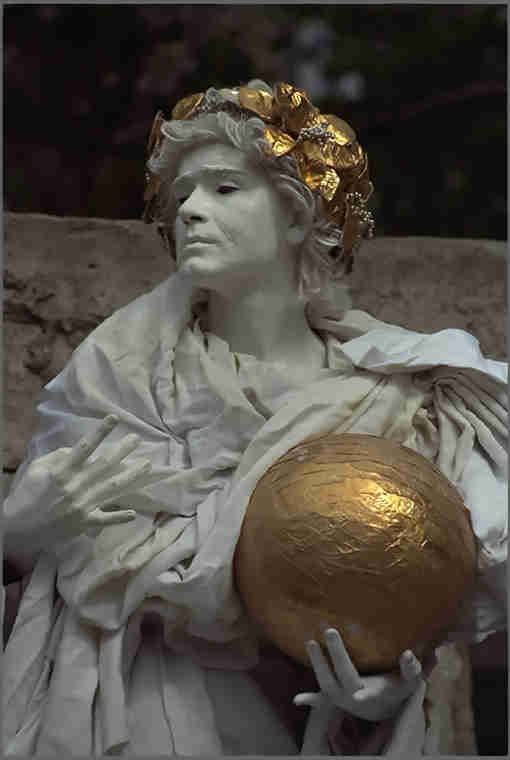}
}
\end{center}
\vspace{-2 mm}
{\caption{Comparison of our pre-editing method with baseline JPEG. The uncompressed input (a) is compressed by JPEG (b), which shows a lot of compression artifacts. We propose to edit the  input image (c) before JPEG compression (d) to obtain a better perceptual quality and lower bit rate. \label{fig:demo}}}
\vspace{-4 mm}
\end{figure}

Far less popular are algorithms that propose to pre-process the image prior to its compression, in order to reduce its entropy, thus avoiding the creation of artifacts in the first place \cite{Oizumi2006preproc,Asshole2005diffusion,Tushabe2007Attribute,dar2018optimized,rott2018deformation,dar2018system}. Indeed, a denoising applied before the compression is often found effective for better encoding performance (e.g. \cite{starck1996Astro}). This line of thinking is scarcer in the literature due to the more complex treatment it induces and the weaker control it provides on the  output artifacts. Still, such a pre-processing approach has a great advantage over the alternatives, as the changes to the image are done on the server side, while the decoder side does not need to be modified nor adjusted. 

In this work we propose to pre-process the image by automatically editing its content, applied before its compression using JPEG standard. Our goal is to modify the image content smartly so as to guarantee that (i) most of the visual information in the image is preserved; (ii) the subsequent compression operates in a much better regime and thus leads to reduced artifacts; and (iii) the edited image after compression is still visually appealing. By considering all these forces holistically, we aim to get creative editing effects that enable the compression-decompression stage to perform at its best for the given bit budget.  
	
While one could pose the proposed editing task as an optimization problem to be solved for each incoming image separately, we take a more challenging route, in which we target the design of a universal deep neural network that performs the required editing on an arbitrary input image. The clear advantage in this approach is the speed with which inference is obtained once the network has been trained. 

Our learning relies on minimizing a loss-function that includes three key penalties, aligned with the above description. The first forces the original and the edited images to be ``sufficiently close'' to each other, while still allowing content editing. A second term penalizes the bit content of the edited image, so as to force the bit-budget constraint while striving for an artifact-free compression. This part is achieved by yet another network \cite{balle2018variational} that predicts the entropy and quality of the image to be compressed. Last, but definitely not least, is a third penalty that encourages the edited image after compression to be visually pleasing. Indeed, our formulation of the problem relies heavily on the availability of a no-reference quality metric, a topic that has seen much progress in recent years \cite{mittal2012noref,xue2013noref,lu2015noref,jin2016noref,mai2016noref,bosse2016noref,Talebi2018noref}. All the above-mentioned ingredients are posed as differentiable machines, enabling an end-to-end effective learning of the  editing operation. An example of the proposed technique is shown in Fig.~\ref{fig:demo}, where the editing operation allows for better perceptual quality and lower bit budget.



\section{Formulating the Problem}
\label{sec:Formulation}

\begin{figure*}[!t]
\vspace{-0 mm}
\begin{center}
\includegraphics*[viewport=1 210 600 410,scale=0.8]{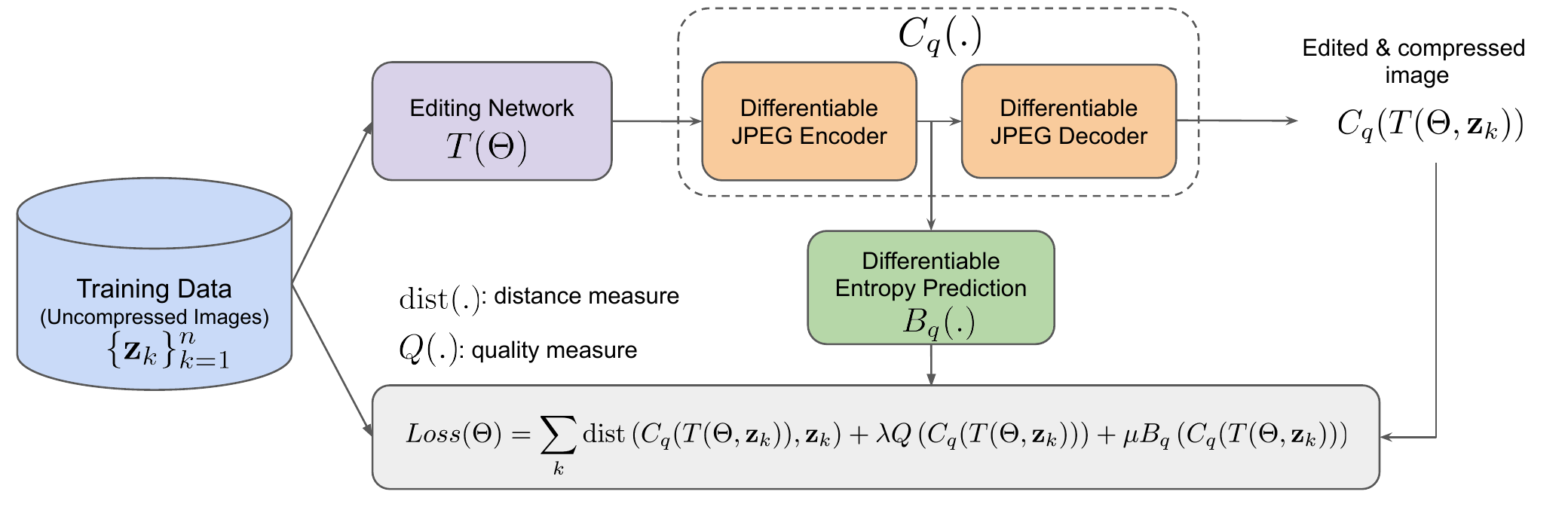}
\end{center}
\vspace{-4 mm}
{\caption{Our learning pipeline for training the image editing $T(\Theta)$. Input image $\textbf{z}_k$ is first edited by our editing network. Then, the edited image is fed to the differentiable JPEG encoder/decoder. The entropy of the quantized DCT coefficients are predicted and used in our training loss. To ensure that the compressed image is close to the uncompressed input, we use a distance measure. We also use a quality term to enforce the human perceptual preference.} \label{fig:training_pipeline}}
\vspace{-4 mm}
\end{figure*}

We start with a few definitions that will help in formulating our problem and its solution. 

\begin{defn}
{\bf (Codec Operation)} We define by $C_R(\x):\mathbb{R}^N \rightarrow \mathbb{R}^N$ the process of compression and decompression of a given image with $R$ bits. This function gets an image $\x$ and produces an image, $C_R(\x)$, possibly with the compression artifacts mentioned above. 
\end{defn}

\begin{defn}
{\bf (Quality Assessment)} We define by $Q(\x):\mathbb{R}^N \rightarrow \mathbb{R}^+$ the process of allocating a no-reference quality to a given image $\x$. The output is a non-negative scalar with values tending to zero for higher quality images.
\end{defn}

\begin{defn} 
{\bf (Distance Measure)} We define by $dist(\x_1,\x_2)$ the distance between two images, $\x_1$ and $\x_2$, of the same size. Our distance function should be defined such that it is ``forgiving'' to minor content changes such as small geometrical shifts or warps,  delicate variations in gray-values, or removal of fine texture. \end{defn}

Armed with the above, we are now ready to formulate our problem. Given an image $\z$ to be compressed with a bit budget of $R$ bits, the common practice is to perform compression and decompression directly, $\hat{\z} = C_R(\z)$, and live with the limitations.

In this work we suggest a novel alternative: We seek a new image $\x$ that is (i) as close as possible to the given image $\z$; (ii) it is compressible using $R$ bits; and most importantly (iii) it is of high quality. Naturally, $\x$ will be an edited variation of $\z$ in which some of the content has been changed, so as to enable good quality despite the compression. Here is our first attempt to formulate this problem:
\begin{eqnarray}\label{eq:formulation1}
\min_{\x} ~dist(\x,\z) + \lambda Q(\x) ~~~s.t.~~ \x=C_R(\x).
\end{eqnarray}
In words, given $\z$ and $R$ we seek an image $\x$ that is close to $\z$, it is of high quality (low value of $Q(\x)$), and it can be represented via $R$ bits. Referring to the constraint, recall that the compression-decompression operation is idempotent, i.e. applying it more than once on a given image results with the same outcome as using it once \cite{joshi2000comparison}. Thus, the constraint aims to say that $\x$ is a feasible outcome of the compression algorithm with the given budget of $R$ bits.


An alternative formulation that may serve the same goal is one in which we fix the quality as a constraint as well,
\begin{eqnarray}\label{eq:formulation2}
\min_{\x} ~dist(\x,\z)  ~~~s.t.~~ \x=C_R(\x)~\mbox{and}~ Q(\x)=Q_0,
\end{eqnarray}
so as to say that whatever happens, we insist on a specific output quality, willing to sacrifice content  accordingly. 

Both problems defined in Equations (\ref{eq:formulation1}) and (\ref{eq:formulation2}), while clearly communicating our goal, are hard to handle. This is mainly due to the non-differentiable nature of the function $C_R(\x)$, and the fact that it is hard to fix a rate $R$ while modifying the image $\x$. While these could be dealt with by a projection point of view (see \cite{Maleki2018compression}), we take a different route and modify our formulation to alleviate these difficulties. This brings us to the following additional definitions: 

\begin{defn}
{\bf (Quality-Driven Codec Operation)} We define by $C_q(\x):\mathbb{R}^N \rightarrow \mathbb{R}^N$ the process of compression and decompression of a given image with a quantization (or quality factor) $q$. This function gets an image $\x$ and produces an image, $C_q(\x)$, possibly with the compression artifacts mentioned above. 
\end{defn}

\begin{defn}
{\bf (Entropy Predictor)} We define by $B_q(\x):\mathbb{R}^N \rightarrow \mathbb{R}^+$ the process of predicting the compression-decompression performance of a specific algorithm (e.g., JPEG) for a given image $\x$ and a quantization level $q$. This function produces the expected file-size (or entropy). 
\end{defn}

Note that by setting $q$, we induce a roughly fixed PSNR on the image after compression-decompression. Thus, by minimizing $B_q(\x)$ with respect to $\x$, we will be aiming for reducing the file size while preserving quality. Returning to our formulation, we add a penalty term, $B_q(\x)$, so as to guarantee that the rate is preserved (or more accurately, controlled). This leads to 
\begin{eqnarray}\label{eq:formulation3}
\min_{\x} ~dist(\x,\z) + \lambda Q(\x) +\mu B_q(\x)~~s.t.~~ \x=C_q(\x).
\end{eqnarray}
The constraint $\x= C_q(\x)$ assures that $\x$ is a valid output of the compression, and it can be alternatively written as\footnote{Admittedly, the notations introduced are a bit cumbersome, as both $B$ and $C$ use the same quantization level $q$. The alternative could have been to divide the compression $C_q$ into an encoder and decoder, and feed the encoder result to $B$ without specifying $q$. We chose to stay with the above formulation for consistency with the opening description.}   
\begin{eqnarray}\label{eq:formulation3a}
\min_{\x} ~dist(C_q(\x),\z) + \lambda Q(C_q(\x)) +\mu B_q(C_q(\x)). 
\end{eqnarray}
If we  have a differentiable proxy for the compression operation $C_q(\cdot)$, the above loss is manageable. 

We could have stopped here, handling this optimization task and getting an edited and compressed image $\x$ for any incoming image $\z$. This could have been a worthy and even fascinating feat by itself, but we leave it for future work. 

As we have already mentioned in the introduction, we aim higher. Our goal is to design a feed-forward CNN that would perform this editing for any given image automatically. Denoting this editing network by $T(\Theta,\z)$, where $\Theta$ are the network parameters to be learned/set, our training loss is given by the following expression:
\begin{eqnarray}\label{eq:formulation4}
Loss(\Theta) & = & \sum_{k} \left[ dist \left(C_q(T(\Theta,\z_k)),\z_k\right) \right. \\
&& \hspace{-0.3in}  \left. + \lambda Q\left(C_q(T(\Theta,\z_k))\right) +\mu B_q\left(C_q(T(\Theta,\z_k))\right) \right]. \nonumber
\end{eqnarray}
This expression simply sums the per-image loss over many training images $\{\z_k\}$, and replaces the edited image $\x_k$ by the network's output $T(\Theta,\z_k)$. Minimizing this loss with respect to $\Theta$, we obtain the editing network, as desired. Our learning pipeline is shown in Fig.~\ref{fig:training_pipeline}.\

		
\section{The Proposed Approach}
\label{sec:Details}

In this section we dive into our implementation of the above-discussed editing idea. We start by specifying the ingredients used within the loss function and then turn to describe the training procedure employed. 


\subsection{Training Loss}

Returning to our definitions, we would like to identify the specific ingredients used in Equation (\ref{eq:formulation4}). In this work we concentrate on the JPEG compression algorithm, due to its massive popularity and the fact that it is central in digital imaging and mobile photography. Our formulation relies on three ingredients:

\noindent {\bf Distance measure:} We seek a definition of $dist(\x_1,\x_2)$ that does not penalize moderate editing changes between the two images. In our implementation we construct this distance measure by feature extraction function, $F(\x):\mathbb{R}^N \rightarrow \mathbb{R}^L$, and use the perceptual loss $\|F(\x_1)-F(\x_2)\|_1$ as our distance \cite{gatys2016perceptual,johnson2016perceptual,ulyanov2016perceptual}. These features could be the activations of an inner layer within the VGG-16 \cite{simonyan2014vgg} or the NIMA \cite{Talebi2018noref} networks,  and  they are  used to  characterize an  image  in  a  domain  that  is  less  sensitive  to  the  allowed perturbations. The  deeper  these  features  are  taken  from, the more daring the editing of the image is expected to be. We experimented with various VGG-16 activations trained for image quality assessment task \cite{Talebi2018noref}, and selected the output of the fifth convolutional block before max pooling as our feature extraction function $F(\x)$.


\noindent {\bf Quality measure:} We assess image quality using  NIMA  \cite{Talebi2018noref}. NIMA is a no-reference image quality assessment machine that has been used for training image enhancement~\cite{talebi2018learned}. 

\noindent {\bf Differentiable JPEG:} As mentioned above, we need to incorporate the function $C_q(\cdot)$ within our loss and thus it should be differentiable. Indeed, as we are working with JPEG, this function does not control the rate but rather the quality factor $q$ when running this compression-decompression. The differentiable encoder/decoder consists of 4 steps:
\begin{enumerate}
\item{Color conversion: We start with an RGB image, and convert to YUV color space. Since color conversion is basically a matrix multiplication, its derivatives are well defined for RGB to YUV and vice versa.}
\item{Chroma downsampling/upsampling: The YUV image may represent full chroma values (YUV444), or subsampled ones (YUV420). The downsampling operation to generate YUV420 is $2\times2$ average pooling. The upsampling is implemented with Bilinear interpolation.}
\item{DCT and inverse DCT: The DCT coefficients ($D_{ij}$) are computed for $8\times8$ image blocks of each YUV channel, separately. Note that the DCT and its inverse are matrix multiplications, and hence differentiable.}
\item{Quantization/Dequantization: Quantization is performed by $8\times8$ tables values $U_{ij}$ as $\lfloor \frac{D_{ij}}{U_{ij}} \rceil$. Note that rounding to the nearest integer operation $\lfloor.\rceil$ has zero derivative almost everywhere, and hence it does not work with our gradient-based learning framework. So, similar to \cite{shin2017jpeg} we employ a third-order polynomial approximation of the rounding operation as $\lfloor x \rceil + (x-\lfloor x \rceil)^3$ where $x$ represents $\frac{D_{ij}}{U_{ij}}$.}
\end{enumerate}

Also, it is worth mentioning that the differentiable JPEG is only used at training, and all the test results presented in the paper use the standard JPEG to measure the bit-rate. So, the differentiable JPEG is only an approximation of the actual JPEG that allows us to train the pre-editing CNN.

\noindent {\bf Entropy prediction:} 
In our framework with JPEG, the discrete entropy of the quantized DCT coefficients should be measured. However, just as described above, the derivatives of the quantization operation are zero almost everywhere, and consequently gradient descent would be ineffective. To allow optimization via stochastic gradient descent, we use the entropy estimator proposed in \cite{balle2018variational}. While \cite{balle2018variational} represents an end-to-end compression scheme, our approach only borrows their entropy estimation technique. Based on this approach, we estimate the bit-rate consumed for JPEG compression by approximating the entropy of the quantized DCT coefficients. The approximated bit-rate can be expressed as $-\mathbb{E}[\log_2 \Pi_{\textbf{d}}]$ where $\textbf{d}$ represents the quantized DCT coefficients. As shown in \cite{Balle17entropy}, $\widehat{\textbf{d}} = \textbf{d} + \Delta \textbf{d}$ is a continuous relaxation of the probability mass function $\Pi_\textbf{d}$, where $\Delta \textbf{d}$ is additive i.i.d. uniform noise with the same width as the quantization bins. This means that the differential entropy of $\widehat{\textbf{d}}$ can be used as an approximation of $\mathbb{E}[\log_2 \Pi_{\textbf{d}}]$. 

We adopt the non-parametric approach of \cite{balle2018variational} to approximate $\Pi_{\widehat{\textbf{d}}}$ marginals. Based on this method, a sigmoid function is used to approximate the cumulative density function of the DCT coefficients $\widehat{\textbf{d}}$. The shift and the scale factors of the sigmoid are learned by a 4-layer neural network. Each intermediate layer of this network consists of 3 neurons followed by $\mbox{tanh}$ activations. The density $\Pi_{\widehat{\textbf{d}}}$ is computed as the derivative of the cumulative function. More implementation details are available in~\cite{tfcompression}.

\begin{figure}[!t]
\vspace{-0 mm}
\begin{center}
\includegraphics*[viewport=140 230 480 530, scale=0.7]{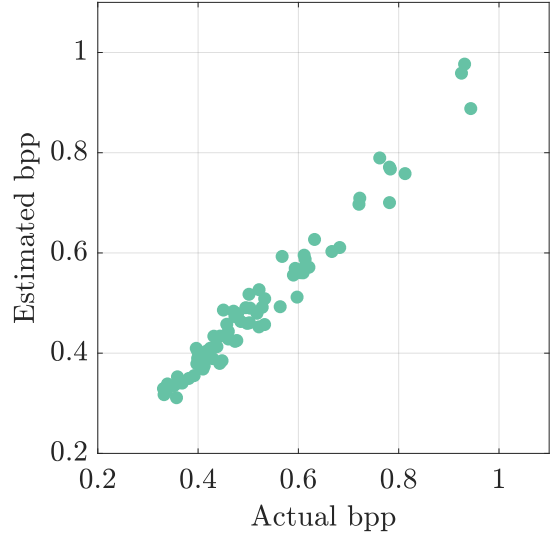}
\end{center}
\vspace{-5 mm}
{\caption{Actual and estimated bpps for Kodak24~\cite{kodak24} images with various JPEG bit-rates. The linear correlation coefficient between the estimated and actual bpps is $0.98$. Each point on this plot corresponds to one of Kodak24 images compressed with a specific JPEG quality factor (total of 72 data points, for 3 quality factors of 10, 15 and 20). Note that these estimated bpps correspond to $\mu B_q(.)$ in the total loss (\ref{eq:formulation4}) with $\mu=0.67$.} \label{fig:bpp}}
\vspace{-4 mm}
\end{figure}

We train separate entropy estimators for each DCT coefficient set obtained from all $8 \times 8$ blocks: (i) Y channel DC (zero-frequency) coefficients, (ii) UV channels DC coefficients, (iii) Y channel non-DC coefficients, (iv) UV channels non-DC coefficients. This approach follows the actual JPEG encoder. The overall entropy is the sum of these four estimated entropies. We also follow the DPCM (Differential Pulse Code Modulation) framework to encode the difference between DC component of the DCT coefficients. Fig. \ref{fig:bpp} shows that the entropy estimator provides a close approximation of the actual JPEG bit-rates. Data points in Fig. \ref{fig:bpp} correspond to a total of 72 images compressed with various JPEG quality factors. Our estimated bit-rates show a strong linear correlation coefficient of $0.98$ with respect to the actual JPEG bit-rates. It is worth noting that as part of our overall training loss, a scalar weight is applied on the estimated bit-rates ($\mu$ in the total loss (\ref{eq:formulation4})). This means that the estimated bit-rats are only required to be accurate up to a scalar factor. Also, these results are computed for the low bit-rate range that we explore in this work (less than $1$ bpp).

\noindent {\bf Total loss:} To summarize, the following is the loss function we use in our experiments: 
\begin{eqnarray} \nonumber
Loss(\Theta) & = & \sum_{k} \left[ dist\left(C_q(T(\Theta,\z_k)),\z_k\right) \right. \\
&& \hspace{-0.3in} \left. + \lambda Q\left(C_q(T(\Theta,\z_k))\right) + \mu B_q\left(C_q(T(\Theta,\z_k))\right)\right] \nonumber
\end{eqnarray}
where the distance function $dist(.)$ represents the perceptual error measures, the image quality $Q(.)$ is computed by NIMA and a total variation measure, and the entropy estimate $B_q(.)$ is computed over the quantized DCT coefficients of the edited image. Note that the same q-factor is applied both in the function $B_q(\cdot)$ and the differentiable JPEG function $C_q(\cdot)$.

\begin{figure}[!t]
\vspace{-0 mm}
\begin{center}
\includegraphics*[viewport=1 1 600 320,scale=0.9]{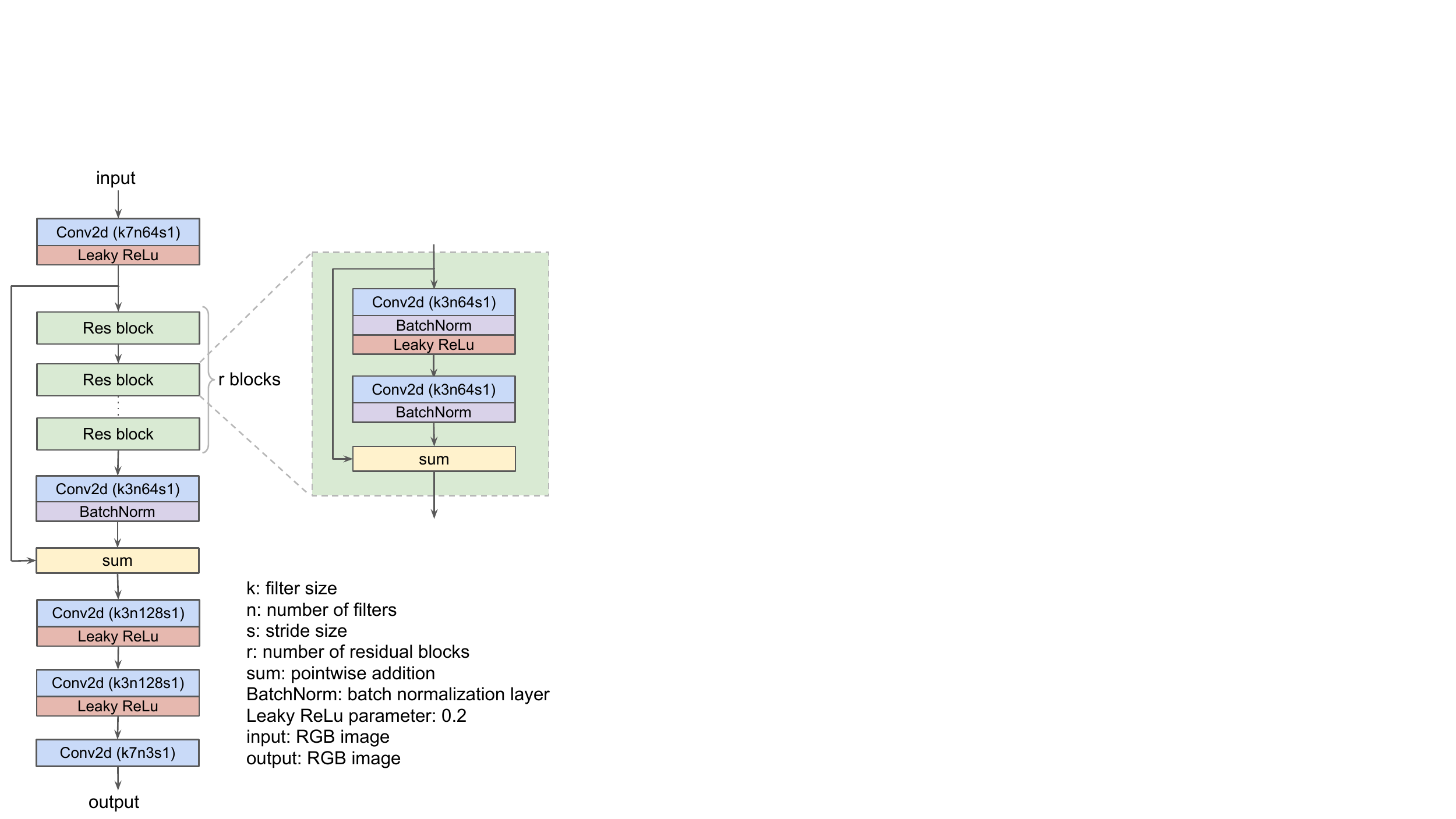}
\end{center}
\vspace{-2 mm}
{\caption{Our image smoothing CNN.} \label{fig:residual_cnn_architecture}}
\vspace{-0 mm}
\end{figure}

\begin{figure*}[!t]
\vspace{-0 mm}
\begin{center}
\includegraphics*[viewport=1 1 600 220,scale=0.8]{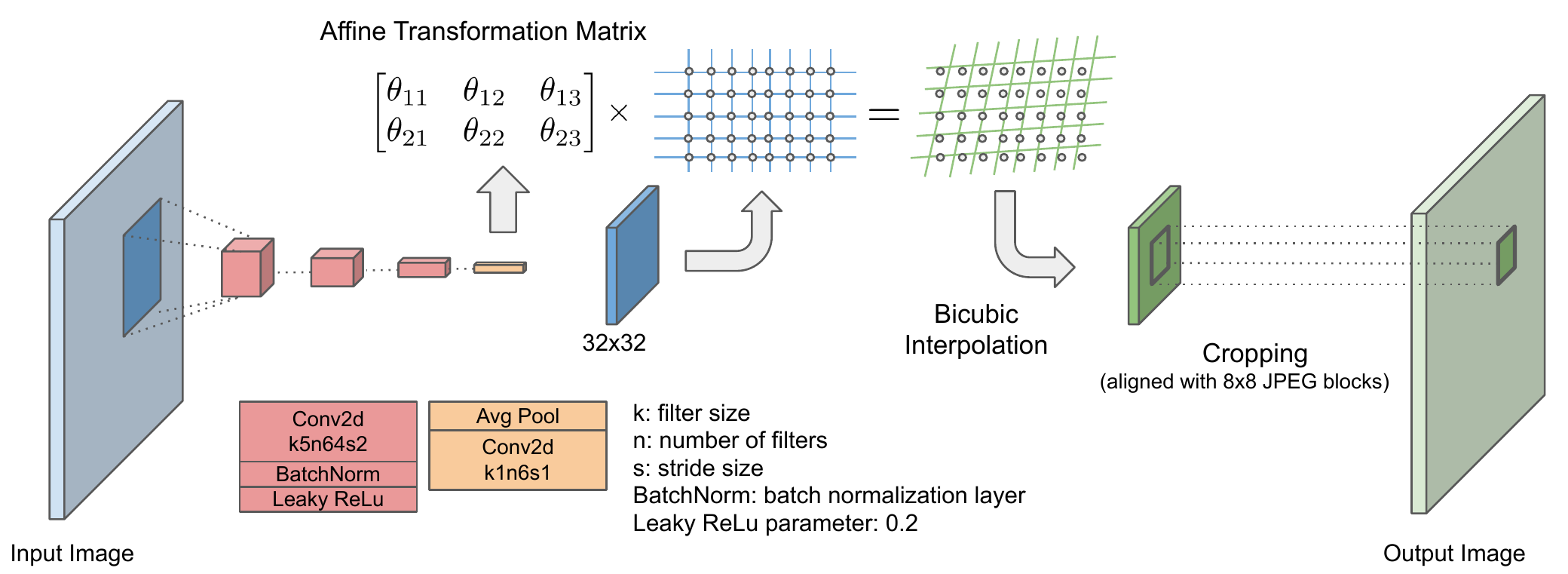}
\end{center}
\vspace{-2 mm}
{\caption{Our patch-based spatial transformer network. The affine transformer parameters of $32\times32$ blocks are obtained from a trainable CNN. Transformed image grid is interpolated to obtain a warped image block of size $32\times32$. Finally, an $8\times8$ central block is extracted.} \label{fig:stn}}
\vspace{0 mm}
\end{figure*}


\begin{figure*}[!t]
\vspace{-0 mm}
\begin{center}
    \includegraphics*[scale=0.21]{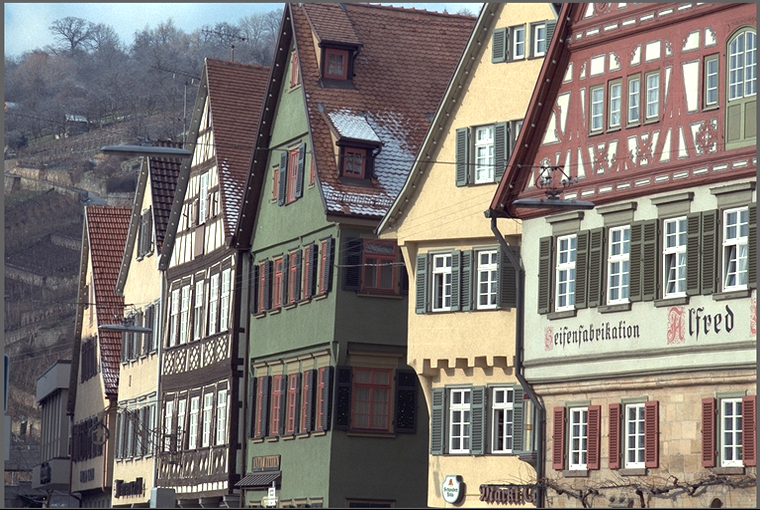}
    \includegraphics*[scale=0.21]{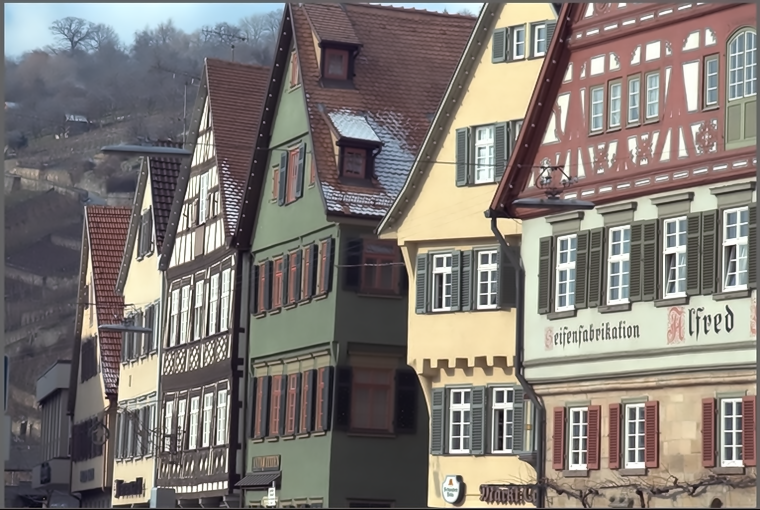}
    \includegraphics*[scale=0.21]{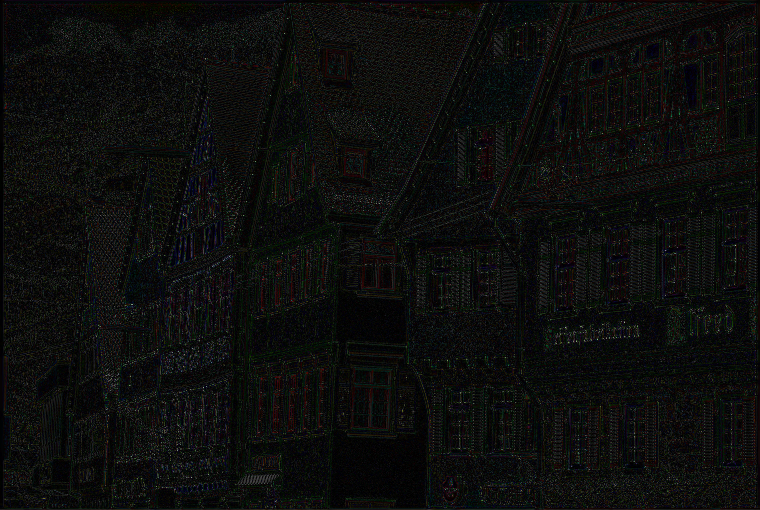}
\\[-1.6mm]
\hspace{-0mm}
\subfigure[\scriptsize Input]{
    \includegraphics*[viewport=180 250 340 340, scale=1.0]{./figures/kodim08.png}
} \hspace{-3.65mm}
\subfigure[\scriptsize Smoothed input]{
    \includegraphics*[viewport=180 250 340 340, scale=1.0]{./figures/kodim08_denoiser_manip_q_20.png}
} \hspace{-3.8mm}
\subfigure[\scriptsize Difference between (a) and (b)]{
    \includegraphics*[viewport=180 250 340 340, scale=1.0]{./figures/kodim08_denoiser_diff_q_20.png}
} \hspace{-0mm}
\end{center}
\vspace{-4 mm}
{\caption{The difference between the input and the smoothed images (without JPEG compression). Our smoothing trained-network removes fine-grain details from the input image to make it more compressible by JPEG. Compressing (a) and (b) images with JPEG encoder at quality factor 20 takes  1.15 and 1.03 bpp, respectively.  \label{fig:diff_smoothing}}}
\vspace{-4 mm}
\end{figure*}

\begin{figure*}[!t]
\vspace{-0 mm}
\begin{center}
    \includegraphics*[scale=0.21]{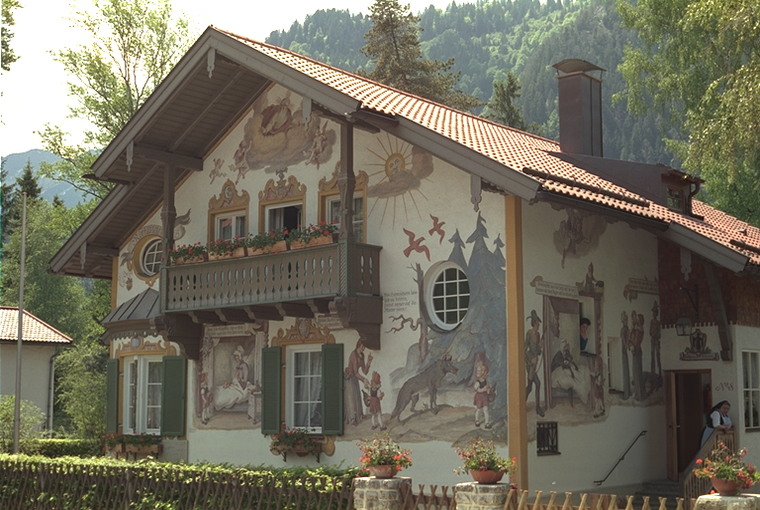}
    \includegraphics*[scale=0.21]{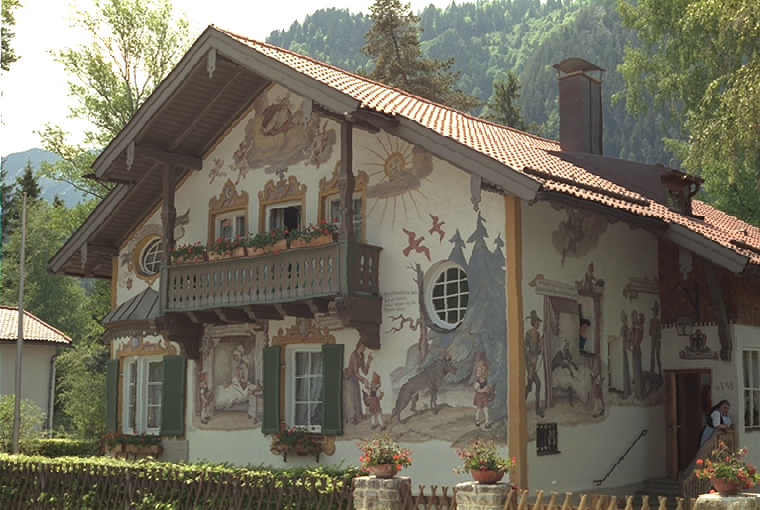}
    \includegraphics*[scale=0.21]{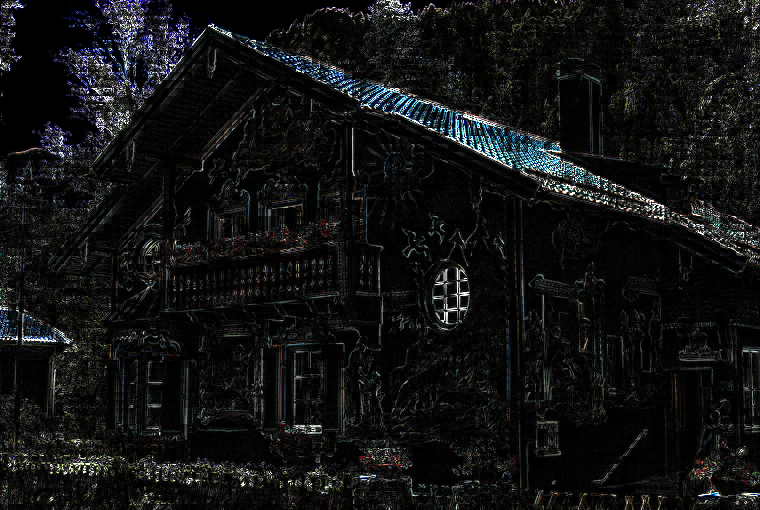}
\\[-1.6mm]
\hspace{-0mm}
\subfigure[\scriptsize Input]{
    \includegraphics*[viewport=180 190 340 280, scale=1.0]{./figures/kodim24.png}
} \hspace{-3.65mm}
\subfigure[\scriptsize Warped input]{
    \includegraphics*[viewport=180 190 340 280, scale=1.0]{./figures/kodim24_stn_manip_q_20.png}
} \hspace{-3.8mm}
\subfigure[\scriptsize Difference of (a) and (b)]{
    \includegraphics*[viewport=180 190 340 280, scale=1.0]{./figures/kodim24_stn_diff_q_20.png}
} \hspace{-0mm}
\end{center}
\vspace{-4 mm}
{\caption{The difference between the input and the warped images (without JPEG compression). Our warping makes spatial transformations on local image patches to make them more compressible by JPEG. Compressing (a) and (b) images with JPEG encoder at quality factor 20 takes  0.725 and 0.708 bpp, respectively.\label{fig:diff_stn}}}
\vspace{-4 mm}
\end{figure*}

\subsection{The Editing Network}

Our editing network consists of two parts; An image smoothing CNN (Fig.~\ref{fig:residual_cnn_architecture}), and a patch-based warping operation (Fig.~\ref{fig:stn}). While the smoothing is similar to a denoiser that controls fine-grained details, the spatial transformer allows for subtle local warps that make the underlying image more compressible~\cite{rott2018deformation}. More details on both parts is given below. 

\subsubsection{The Smoothing Network}
\label{sec:smoothing}

Our image smoothing CNN is shown in Fig.~\ref{fig:residual_cnn_architecture}. This convolutional neural network is similar to the residual CNN of Ledig et al.~\cite{ledig2017photo}. This architecture has $r=2$ identical residual blocks, with 3$\times$3 kernels and 64 feature maps followed by batch normalization~\cite{ioffe2015batch} layers and Leaky ReLu (instead of parametric ones)~\cite{glorot2011deep} activations. 
To avoid boundary artifacts, the input image and feature maps are symmetrically padded before convolutions. We also append a smoothing strength factor (noise standard deviation) and the JPEG quality factor as extra channels to the input image. We observed that this additional information helps with better generalization of the model. 

Examples of using the trained smoothing network are shown in Fig.~\ref{fig:diff_smoothing}. These images are not compressed by JPEG, and only represent edits applied to the input. The difference image shows that our editing removes fine details. Note that compressing the smoothed image with JPEG encoder at quality factor 20 takes 1.03 bpp, whereas the same encoder takes 1.15 bpp for compressing the input image. 

\subsubsection{The Spatial Transformer Network (STN)}
\label{sec:stn}

As shown by Rott et al.~\cite{rott2018deformation}, local image deformations can lead to DCT domain sparsity and consequently better compressibility. Unlike~\cite{rott2018deformation} that solves an alternating optimization with an optical flow, we use the differentiable spatial transformer network~\cite{jaderberg2015spatial}. STN learns 6 parameters for an affine local transformation that allows cropping, translation, rotation, scale, and skew to be applied on the input (Fig.~\ref{fig:stn}). We apply STN on overlapping blocks of size $32\times32$, and then we extract central crops of size $8\times8$ that are aligned with JPEG blocks. Since each $32\times32$ block is warped separately, this can cause inconsistency near the boundary of cropped blocks. To alleviate this, all overlapped grid values are averaged across neighboring blocks.  

Examples of using the trained STN are shown in Fig.~\ref{fig:diff_stn}. The STN warps textures and edges locally to make the $8\times8$ blocks more compressible by JPEG encoder. Compressing the input and deformed images in Fig.~\ref{fig:diff_stn}(a) and Fig.~\ref{fig:diff_stn}(b) with JPEG encoder at quality factor 20 requires 0.725 bpp and 0.708 bpp, respectively.

To take advantage of both editing stages, we cascade the smoothing and warping operations. While the smoothing allows for less blockiness artifacts, the STN leads to texture preservation. Next, we discuss our training data.

\subsection{Data}

Our editing networks are trained on uncompressed images. To this end, we use burst processed images of Hasinoff et al.~\cite{hasinoff2016burst}, which provides 3640 images of size 12 mega pixels. All images are converted to 8-bit PNG format. We extract about 120K non-overlapping patches of size $480 \times 640$, and use them to train our model. We also create a test set with $10\%$ of the data.


\section{Relation to Prior Work}
\label{sec:Prior}

We pause our main story for a while and discuss the rich literature on combating compression artifacts. Our goal is to give better context to the suggested methodology by presenting the main existing alternatives. Note, however, that this does not constitutes an exhaustive scan of the existing literature, as this is beyond the scope of this work. We survey these algorithms by dividing them into categories based on their core strategies: 

\noindent {\bf Post-Processing Algorithms}  \cite{zakhor1992iterative,shen1998review,chen2001adaptive,triantaffilidis2002blockiness,alter2005adapted,averbuch2005deblocking,kartalov2007adaptive,weiss2008compression,zhai2009efficient,jung2012image,zhang2013compression,dar2016postprocessing,du2007post,du2011new,kwon2015efficient}: Those are the most common methods available, operating on the image after the compression-decompression damage has already been induced. Algorithms of this sort that are designed in the context of the JPEG format are known as deblocking algorithms. The idea behind these methods, be it for JPEG or any other transform-based coder, is quite simple, even though there are many ways to practice it; Given the compressed-decompressed image and knowing the quantization levels and the transform applied, the original image to be recovered must lie in a convex set that has a rotated hyper-rectangle shape. A recovery algorithm should seek for the most probable image within this set, something that could be done by relying on various regularization strategies. While some algorithms make use of this very rationale directly, others relax it in various ways, by simplifying the constraint set to a sphere, by forcing the recovery algorithm to take a specific shape, and more. At it simplest form, such a deblocking could be a simple linear filter applied to the boundaries between adjacent blocks.    

\noindent {\bf Deep-Learning Based Solutions}  \cite{Dong2015deep,Wang2016deep,Galteri2017deep,Cavigelli2017deep}: Still under the regime of post-processing, recent solutions rely on deep neural networks, trained in a supervised fashion to achieve their cleaning goal. These methods tend to be better performing, as their design targets the recovery error directly, instead of relying on model-based restoration methods.  
    
\noindent {\bf Scale-Down and Scale-Up} \cite{bruckstein2003down,tsaig2005variable,lin2006adaptive,wu2009low}: An entirely different way to avoid compression artifacts is to scale-down the image before compression, apply the compression-decompression on the resulting smaller image, and scale-up the outcome in the client after decompression. This approach is especially helpful in low bit-rates, since the number of blocks is reduced, the bit stream overhead reduce along with it, and the scale-up at the client brings an extra smoothing. Variations over this core scheme have been proposed over the years, in which the scale-down or up are optimized for better end-to-end performance.  
    
\noindent {\bf Pre-Processing Algorithms} \cite{Oizumi2006preproc,Asshole2005diffusion,Tushabe2007Attribute,starck1996Astro}: It is well known that compression-decompression often behaves as a denoiser, removing small and faint details from the image. Nevertheless, applying a well-designed denoiser prior to the compression may improve the overall encoding performance by better prioritizing the content to be treated. The existing publications offering this strategy have typically relied on this intuition, without an ability to systematically design the pre-filter for best end-to-end performance, as the formulation of this problem is quite challenging.   

\noindent {\bf Deformation Aware Compression}
\cite{rott2018deformation}: While this work offers a pre-processing of the image along the same lines as described above, we consider it as a class of its own because of two reasons: (i) Rather than using a denoiser, the pre-process applied in this work is a geometrical warp, which re-positions elements in the image to better match the coder transform and block-division; and (ii)  the design of the warp is obtained by an end-to-end approximate optimization method. Indeed, this paper has been the source of inspiration behind our ideas in this work. 

Our proposed method joins the list of pre-processing based artifact removal algorithms, generalizing the work in \cite{rott2018deformation} in various important ways: (i) our method could accommodate more general editing effects; (ii) its application is simple and fast, once the editing network has been trained; and (iii) we employ a no-reference image quality assessment that supports better quality outcomes. As already mentioned, the pre-processing strategy has a unique advantage over the alternative methods in the fact that the decoder does not have to be aware of the manipulations that the image has gone through, applying a plain decoding, while leaving the burden of the computations to the encoder. That being said, we should add that this approach can be easily augmented with a post-processing stage, for fine-tuning and improving the results further. 

We conclude this survey of the relevant literature by referring to two recent and very impressive papers. The work reported in \cite{blau2019perceptual} offers a theoretical extension of the classic rate-distortion theory by incorporating the perceptual quality of the decompressed-image, exposing an unavoidable trade-off between distortion and visual quality. Our work practices this very rationale by sacrificing image content (via pre-editing) for obtaining better looking compressed-decompressed images. The work by Agustsson et. al \cite{agustsson2019perceptual} offers a GAN-based learned compression algorithm that practically trades visual quality for distortion. While aiming for the same goal as our work, \cite{agustsson2019perceptual} replaces the whole compression-decompression process, whereas we insist on boosting available standard algorithms, such as JPEG, due to their massive availability and spread use.  


\section{Experimental Results}
\label{sec:Experiments}

\begin{figure}[!t]
\vspace{-1 mm}
\begin{center}
\includegraphics*[viewport=100 230 480 560, scale=0.6]{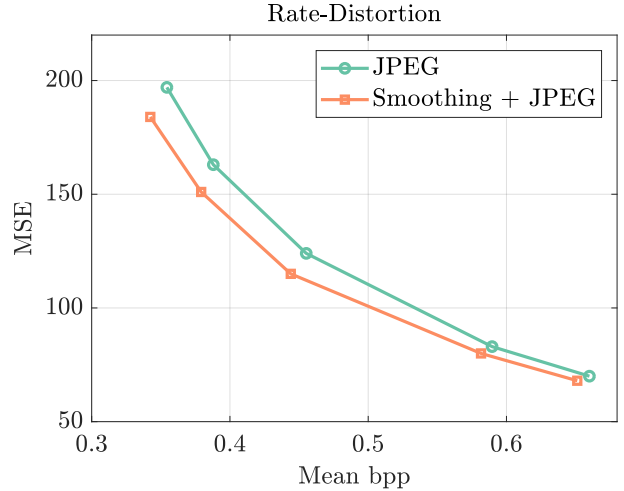}
\end{center}
\vspace{-5 mm}
{\caption{MSE vs. mean bit-rate for the Kodak dataset~\cite{kodak24}.} \label{fig:rate_distortion}}
\vspace{-4 mm}
\end{figure}

\begin{figure*}[!t]
\vspace{-2 mm}
\begin{center}
    \includegraphics*[scale=0.13]{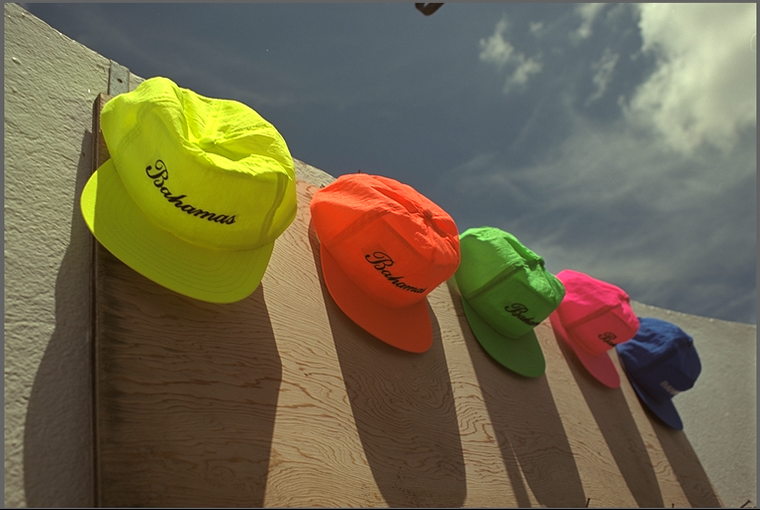}
    \includegraphics*[scale=0.13]{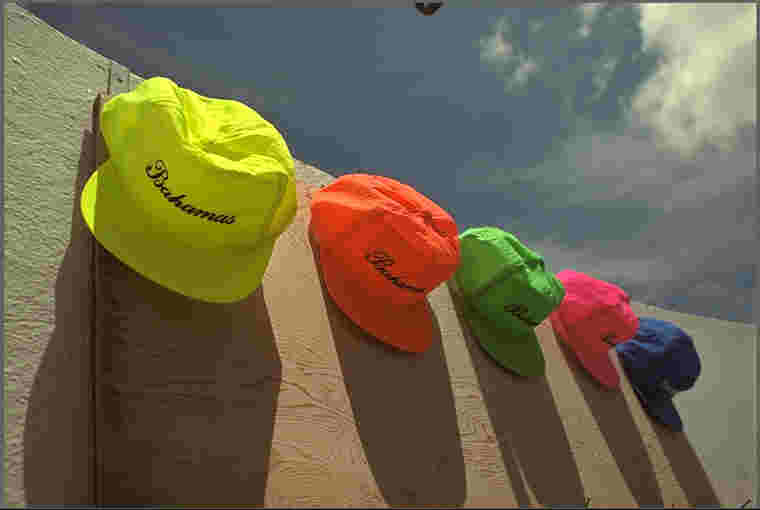}
    \includegraphics*[scale=0.13]{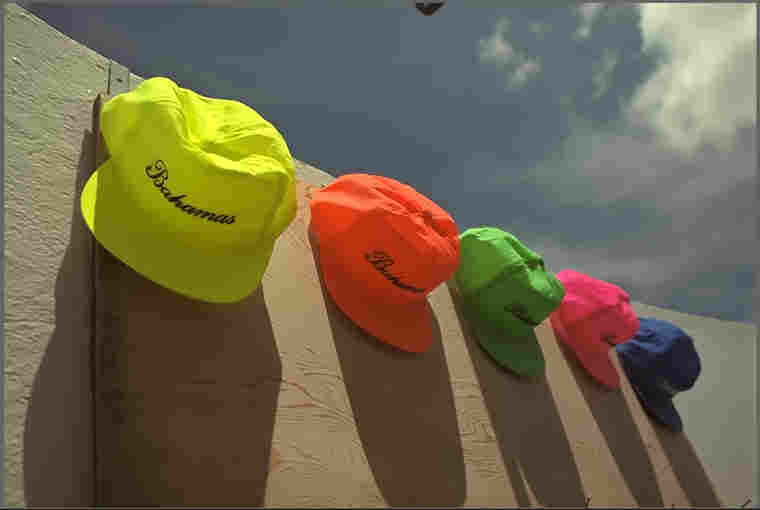}
    \includegraphics*[scale=0.13]{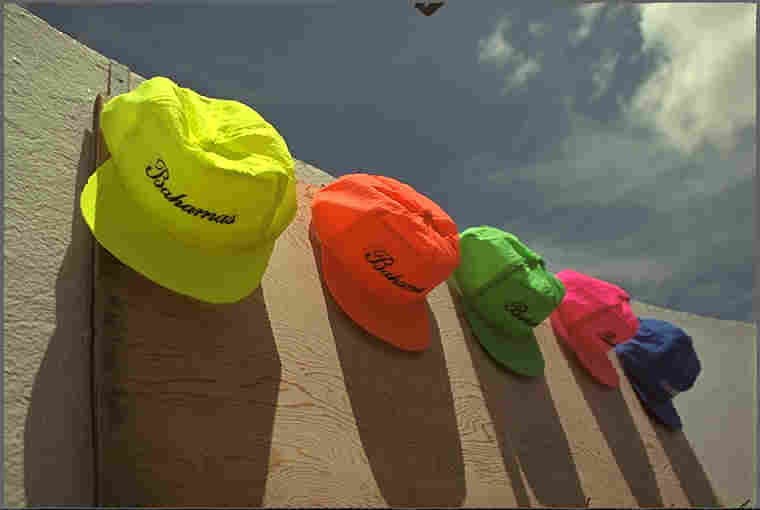}
    \includegraphics*[scale=0.13]{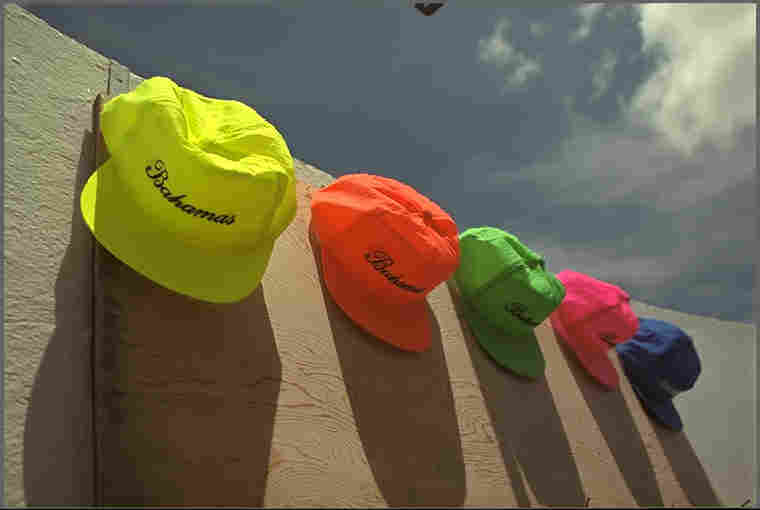}
\\[-1.6mm]
\hspace{-1.2mm}
\subfigure[{\fontsize{6.5pt}{1pt}\selectfont Input}]{
    \includegraphics*[viewport=380 20 540 120, scale=0.615]{./figures/kodim03.png}
} \hspace{-3.50mm}
\subfigure[{\fontsize{6.5pt}{1pt}\selectfont Baseline JPEG (0.3529 bpp)}]{
    \includegraphics*[viewport=380 20 540 120, scale=0.615]{./figures/kodim03_denoiser_comp_baseline_q_15.jpg}
} \hspace{-3.30mm}
\subfigure[{\fontsize{6.5pt}{1pt}\selectfont Smoothing+JPEG (0.3508 bpp)}]{
    \includegraphics*[viewport=380 20 540 120, scale=0.615]{./figures/kodim03_denoiser_comp_manip_q_15.jpg}
} \hspace{-3.6mm}
\subfigure[{\fontsize{6.5pt}{1pt}\selectfont STN+JPEG (0.3525 bpp)}]{
    \includegraphics*[viewport=380 20 540 120, scale=0.615]{./figures/kodim03_stn_comp_manip_q_15.jpg}
} \hspace{-3.7mm}
\subfigure[{\fontsize{6.2pt}{1pt}\selectfont STN+Smoothing+JPEG (0.3495 bpp)}]{
    \includegraphics*[viewport=380 20 540 120, scale=0.615]{./figures/kodim03_hybrid_comp_manip_q_15.jpg}
} \hspace{-0mm}

\end{center}
\vspace{-4 mm}
{\caption{Compression performance with our proposed framework. (c) Smoothing the image before compression leads to less blockiness and color artifacts (0.6\% bpp reduction), (d) The STN network generates more compressible details, (e) The combination of smoothing and STN lowers the bit-rate by 0.9\% bpp. Our perceptual study shows that (e) is the most preferred image. \label{fig:comp_smoothing}}}
\vspace{-1 mm}
\end{figure*}

\begin{figure*}[!t]
\vspace{-2 mm}
\begin{center}
    \includegraphics*[scale=0.13]{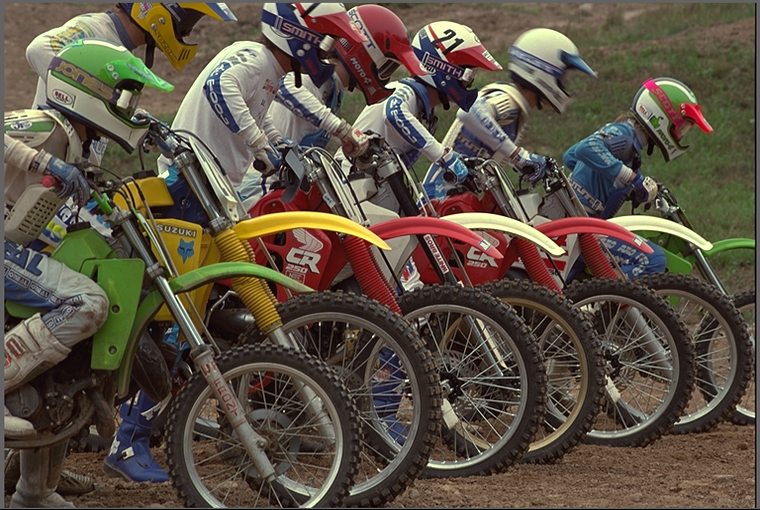}
    \includegraphics*[scale=0.13]{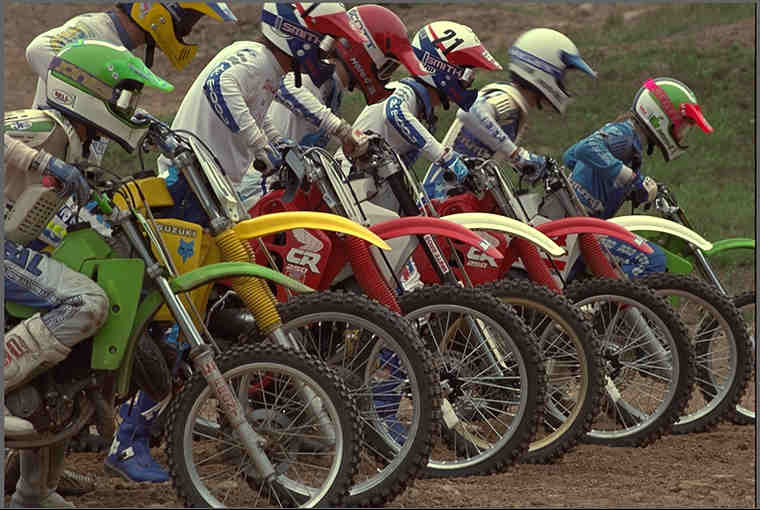}
    \includegraphics*[scale=0.13]{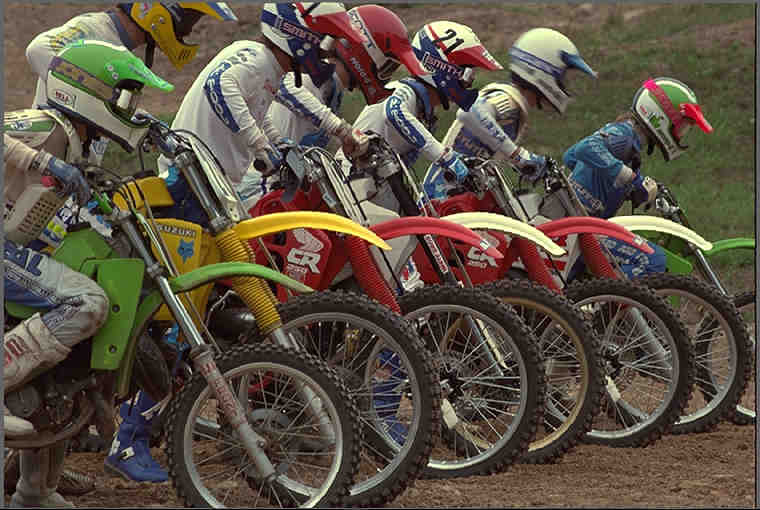}
    \includegraphics*[scale=0.13]{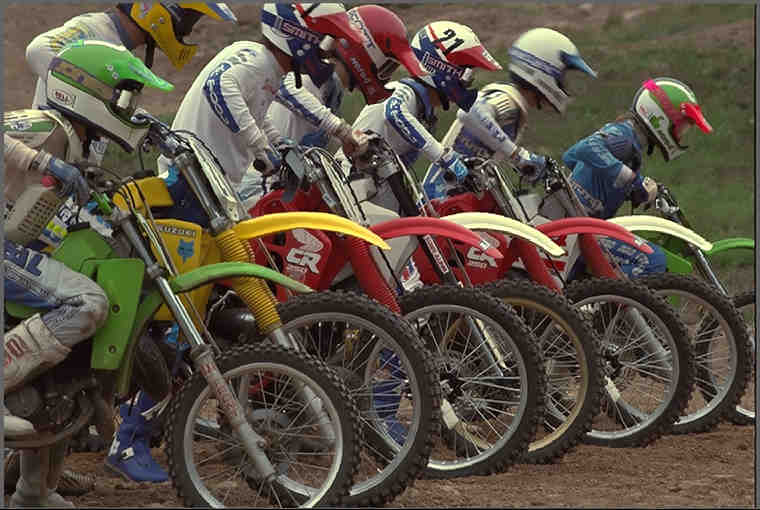}
    \includegraphics*[scale=0.13]{./figures/kodim05_hybrid_comp_manip_q_25.jpg}
\\[-1.6mm]
\hspace{-1.2mm}
\subfigure[{\fontsize{6.5pt}{1pt}\selectfont Input}]{
    \includegraphics*[viewport=300 180 460 270, scale=0.615]{./figures/kodim05.png}
} \hspace{-3.50mm}
\subfigure[{\fontsize{6.5pt}{1pt}\selectfont Baseline JPEG (1.0567 bpp)}]{
    \includegraphics*[viewport=300 180 460 270, scale=0.615]{./figures/kodim05_stn_comp_baseline_q_25.jpg}
} \hspace{-3.30mm}
\subfigure[{\fontsize{6.5pt}{1pt}\selectfont Smoothing+JPEG (1.0423 bpp)}]{
    \includegraphics*[viewport=300 180 460 270, scale=0.615]{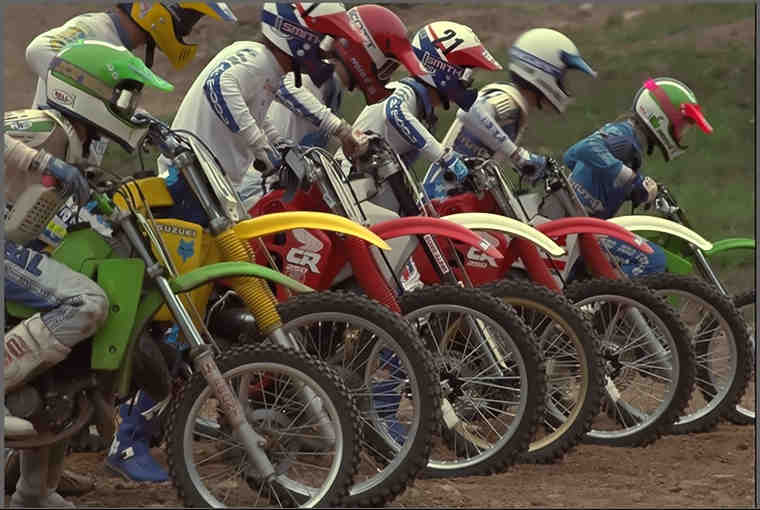}
} \hspace{-3.6mm}
\subfigure[{\fontsize{6.5pt}{1pt}\selectfont STN+JPEG (1.0485 bpp)}]{
    \includegraphics*[viewport=300 180 460 270, scale=0.615]{./figures/kodim05_stn_comp_manip_q_25.jpg}
} \hspace{-3.7mm}
\subfigure[{\fontsize{6.2pt}{1pt}\selectfont STN+Smoothing+JPEG (1.0445 bpp)}]{
    \includegraphics*[viewport=300 180 460 270, scale=0.615]{./figures/kodim05_hybrid_comp_manip_q_25.jpg}
} \hspace{-0mm}

\end{center}
\vspace{-4 mm}
{\caption{Compression performance with our proposed framework. (c) Smoothing the image before compression leads to less blockiness and less details (1.3\% bpp reduction), (d) The STN network generates more compressible details at lower bit-rate compared to the baseline (0.8\% bpp reduction), (e) The combination of smoothing and STN lowers the bit-rate by 1.1\% bpp. Our perceptual study shows that (d) is the most preferred image. \label{fig:comp_stn}}}
\vspace{-1 mm}
\end{figure*}

\begin{figure}[!t]
\vspace{-0 mm}
\begin{center}
\includegraphics*[viewport=100 230 480 540, scale=0.6]{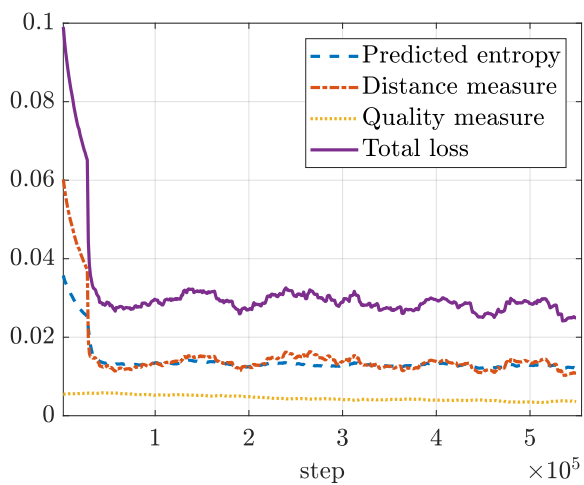}
\end{center}
\vspace{-5 mm}
{\caption{Our training loss components during gradient descent with JPEG quality factor in the range [8,25]. For better display, all losses are smoothed.} \label{fig:loss}}
\vspace{-4 mm}
\end{figure}

\begin{figure*}[!t]
\vspace{-0 mm}
\begin{center}
    \includegraphics*[scale=0.21]{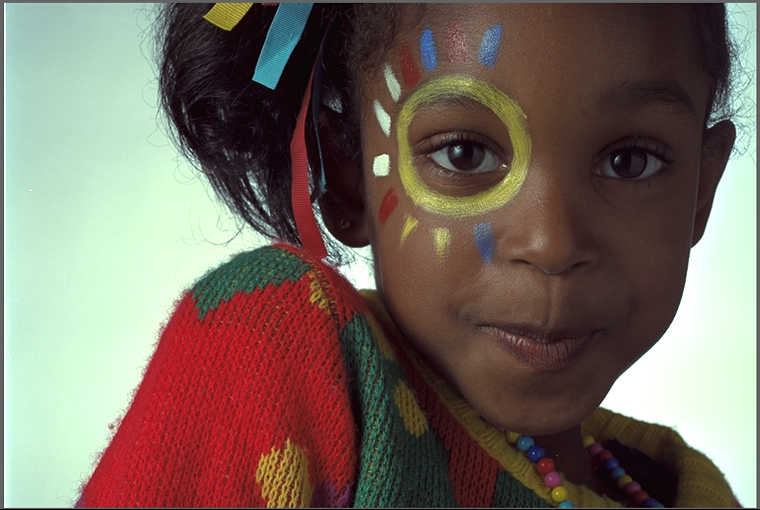}
    \includegraphics*[scale=0.21]{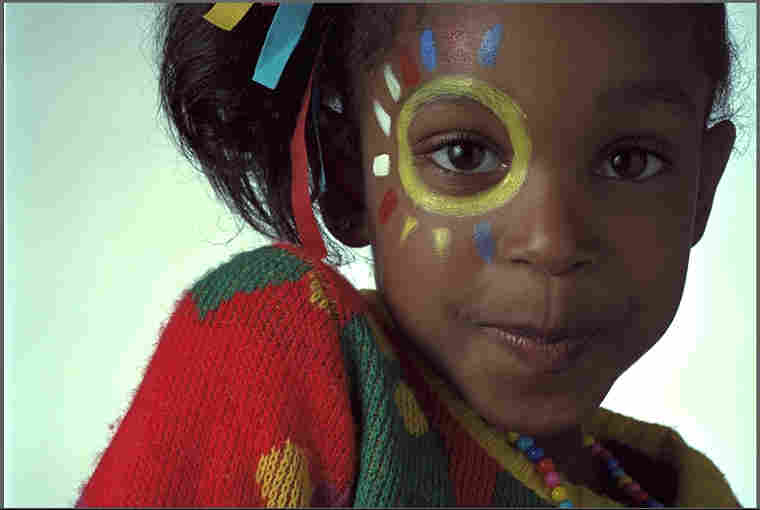}
    \includegraphics*[scale=0.21]{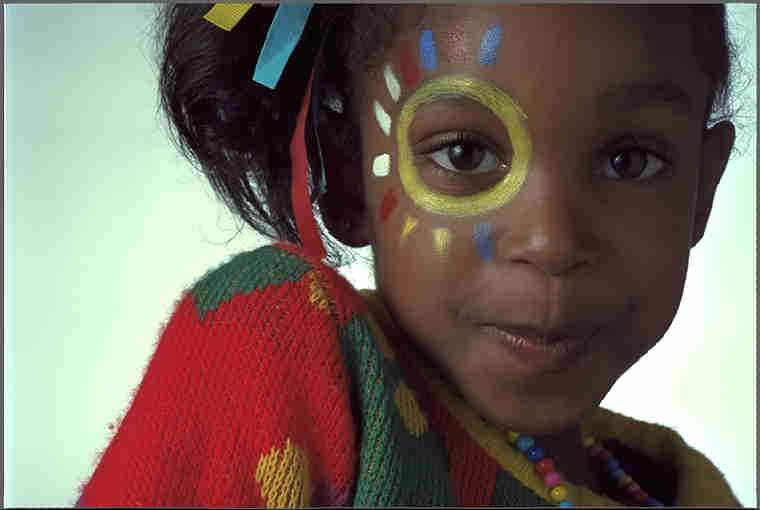}
\\[-1.6mm]
\hspace{-0mm}
\subfigure[\scriptsize Input]{
    \includegraphics*[viewport=520 230 680 370, scale=1.0]{./figures/kodim15.png}
} \hspace{-3.65mm}
\subfigure[\scriptsize Baseline JPEG (0.4705 bpp)]{
    \includegraphics*[viewport=520 230 680 370, scale=1.0]{./figures/kodim15_hybrid_comp_baseline_q_20.jpg}
} \hspace{-3.8mm}
\subfigure[\scriptsize Smoothing + STN + JPEG (0.4355 bpp)]{
    \includegraphics*[viewport=520 230 680 370, scale=1.0]{./figures/kodim15_hybrid_comp_manip_q_20.jpg}
} \hspace{-0mm}
\end{center}
\vspace{-4 mm}
{\caption{Compression performance for applying smoothing and STN (7.4\% bpp reduction). \label{fig:comp_hybrid_1}}}
\vspace{-0 mm}
\end{figure*}

\begin{figure*}[!t]
\vspace{-0 mm}
\begin{center}
    \includegraphics*[scale=0.275]{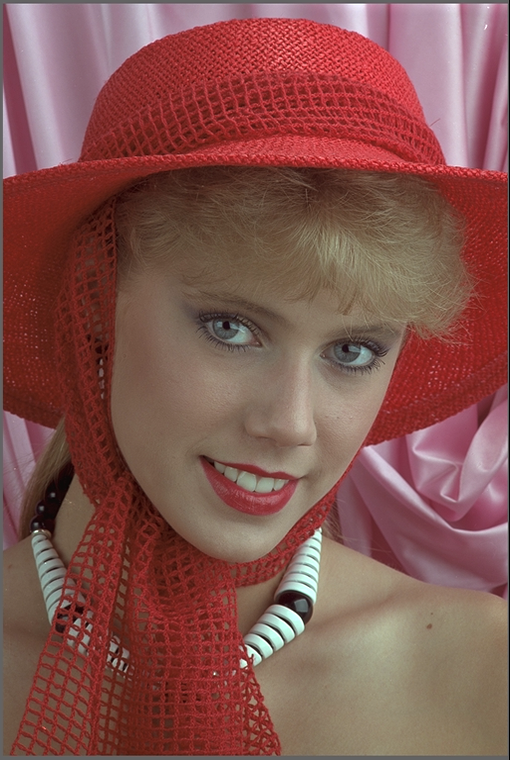}
    \includegraphics*[scale=0.275]{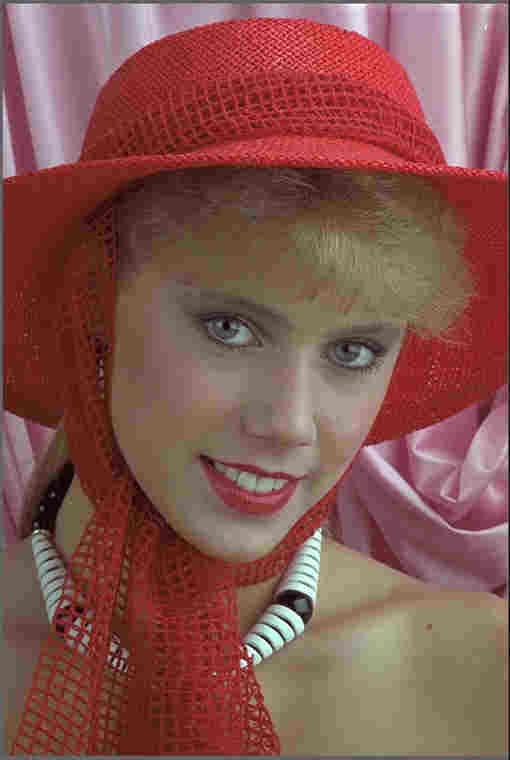}
    \includegraphics*[scale=0.275]{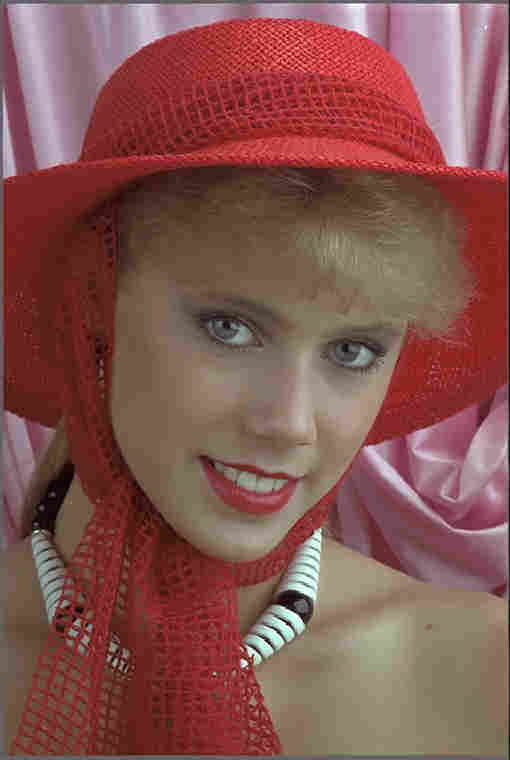}
\\[-1.6mm]
\hspace{-0mm}
\subfigure[\scriptsize Input]{
    \includegraphics*[viewport=110 320 250 480, scale=1.0]{./figures/kodim04.png}
} \hspace{-3.65mm}
\subfigure[\scriptsize Baseline JPEG (0.4293 bpp)]{
    \includegraphics*[viewport=110 320 250 480, scale=1.0]{./figures/kodim04_hybrid_comp_baseline_q_15.jpg}
} \hspace{-3.8mm}
\subfigure[\scriptsize Smoothing + STN + JPEG (0.4169 bpp)]{
    \includegraphics*[viewport=110 320 250 480, scale=1.0]{./figures/kodim04_hybrid_comp_manip_q_15.jpg}
} \hspace{-0mm}
\end{center}
\vspace{-4 mm}
{\caption{Compression performance for applying smoothing and STN (2.9\% bpp reduction). \label{fig:comp_hybrid_2}}}
\vspace{-4 mm}
\end{figure*}

\begin{figure}[!t]
\vspace{-0 mm}
\begin{center}
\includegraphics*[viewport=100 230 480 560, scale=0.6]{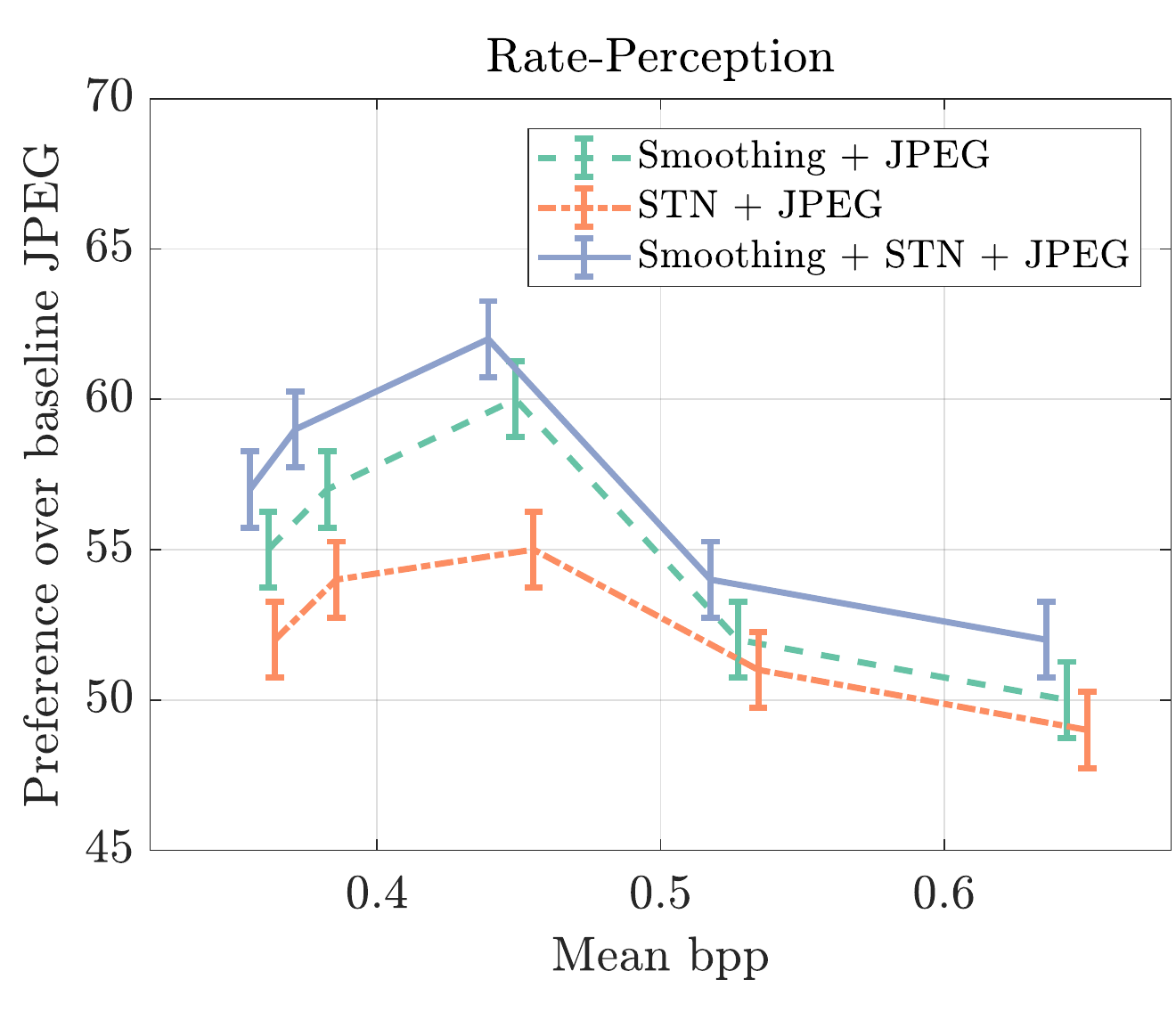}
\end{center}
\vspace{-5 mm}
{\caption{Percentage of human raters preference for pairwise comparison between our result and baseline JPEG. Each data point is an average of 480 ratings (24 Kodak images~\cite{kodak24} and 20 human raters).} \label{fig:rate_perception}}
\vspace{-4 mm}
\end{figure}

In this section our results are discussed and compared to other methods. Our train and test are performed on a single Nvidia GPU V100 with 16GB RAM. At training, images are cropped to $480 \times 640$, and testing is performed on the Kodak dataset~\cite{kodak24}. We use the Adam optimizer \cite{kingma2014adam} with learning rate set to $0.0001$, and batch size as 1. The editing notworks are trained for $5 \times 10^5$ steps of stochastic gradient descent. Weights from the NIMA are kept fixed during training.

In order to train the STN and the smoothing network, we randomly sample the JPEG quality factor from a uniform distribution in the range $[8,25]$ at each step of the gradient descent. This allows our editing to be effective for a range of bit-rates. At test time, we compare our results with the baseline JPEG at comparable bit-rates. To compress a test image at various bit-rates, we adjust the JPEG quality factor to ensure that our result compresses with the closest fewer bits.

We trained the smoother and STN networks separately, and then cascaded and fine-tuned them jointly. Next we discuss each model.

\noindent {\bf The Smoothing Network}: The smoothing model is trained with an $L_2$ distance measure $dist(\x_1,\x_2)=\|\x_1-\x_2\|_2^2$. The weights in our loss (Eq. \ref{eq:formulation4}) are set as $\lambda=0$ and $\mu=0.01$. Note that this weight selection allows for optimizing the typical rate-MSE curve. Training images are augmented with random additive white Gaussian noise to enforce smoothing property in the resulting network. We randomly vary the standard deviation of the noise in the range $[0,0.15]$ at each training step, and append the noise standard deviation and JPEG quality factor as extra channels to the input RGB image. Rate-distortion curves of the regular JPEG and the smoother content are shown in Fig. \ref{fig:rate_distortion}. As expected, the smoothing improves upon the baseline JPEG. Note that these results are obtained before fine-tuning the smoother with the STN. Examples of the smoother editing are shown in Fig. \ref{fig:comp_smoothing}, where color degradation and blockiness artifacts are more visible in the baseline JPEG, compared to our results.

\noindent {\bf The Spatial Transformer Network}: The STN model is obtained by training with the distance measure $dist(\x_1,\x_2)=\|F(\x_1)-F(\x_2)\|_1$, where $F(.)$ represents the NIMA activations. This is due to the fact that $L_1$ or $L_2$ loss does not allow spatial transformations and warps. In contrast, NIMA CNN activations~\cite{Talebi2018noref} are relatively tolerant to such transformations. The weights in our training loss (Eq. \ref{eq:formulation4}) are set as $\lambda=0.02$ and $\mu=1.0$. Our results for the STN network are shown in Fig. \ref{fig:comp_stn}. Our editing of the input images allows to preserve structures and textures more effectively. The local deformations of the STN seem to make certain image textures more compressible. Note that this is a different behavior than the smoother's effect. Also, it is worth noting that the spatial transformations can significantly reduce PSNR, and consequently rate-distortion analysis is not a fair performance evaluation of the STN.

\noindent {\bf The Cascaded Model}: Our weighted losses for the cascaded smoother and STN are shown in Fig. \ref{fig:loss}. The weights in the training loss are similar to the weights used in the STN model. Our experiments suggest that the weighted predicted entropy and the distance measure should be close to each other. Also, as discussed in \cite{talebi2018learned}, the quality measure is most effective when its contribution is limited to a fraction of the total loss. We fine-tune both the smoother and STN networks jointly and present the results in Figs. \ref{fig:comp_hybrid_1} and \ref{fig:comp_hybrid_2}. The cascade editor seems to present comparable details to the baseline, but with less JPEG artifacts.

We carried a human evaluation study to compare our proposed framework with baseline JPEG. We used Amazon Mechanical Turk with pairwise comparison for this task. We asked raters to select the image with better quality. We processed 24 images from Kodak dataset~\cite{kodak24} with our smoothing and warping (STN) frameworks and compared them with their baseline JPEG counterparts at similar bit-rate. Comparisons are made by 20 human raters, and average percentage of the raters preference over baseline JPEG is reported in Fig.~\ref{fig:rate_perception}. As can be seen, both STN and our smoothing show perceptual preference of more than $50\%$ for bit-rates smaller than 0.5 bpp. For higher bit-rates our methods did not provide a statistically significant advantage over baseline. Also, we observed that smoothing consistently outperforms STN.    

\begin{figure}[!t]
\vspace{-0 mm}
\begin{center}
\includegraphics*[viewport=100 230 480 560, scale=0.6]{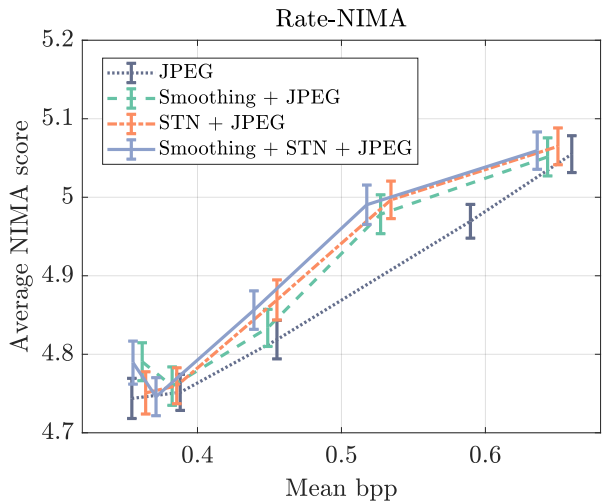}
\end{center}
\vspace{-5 mm}
{\caption{Comparing our results with respect to the NIMA score~\cite{Talebi2018noref}. These scores range from 1 to 10, with 10 indicating the highest quality. Each data point is an average of scores from non-overlapping crops extracted from Kodak images.} \label{fig:rate_nima}}
\vspace{-4 mm}
\end{figure}

We also compare our results with respect to the NIMA score~\cite{Talebi2018noref} in Fig.~\ref{fig:rate_nima}. We computed NIMA scores for non-overlapping patches extracted from Kodak dataset and averaged the resulting scores. Fig.~\ref{fig:rate_nima} indicates that the proposed pre-editing shows an advantage over the baseline JPEG for bpp $\in[0.45,0.6]$. For bit-rates outside this range, the perceptual advantage over the baseline JPEG disappears.

\noindent {\bf Ablation Study}: The impact of our design choices such as the perceptual loss and the model depth are discussed in the following. More explicitly, the CNN activations $F(.)$ in the perceptual loss $\|F(\x_1)-F(\x_2)\|_1$ are intermediate VGG-16 layers trained for quality assessment~\cite{Talebi2018noref}. We tried the third, the fourth, and the fifth convolutional blocks as our feature functions to train various STN models. We observed that the average NIMA score obtained from the STN model trained with the fifth layer of VGG is $2.3\%$ better than the model trained with the fourth layer activations. This gap increases to $4.0\%$ when comparing STN models trained with the fifth and the third layer activations. These measurements are made for nearly comparable bit-rates.

Another important parameter in our framework is the depth of the smoothing CNN shown in Fig.~\ref{fig:residual_cnn_architecture}. The depth parameter is effectively controlled by the number of residual blocks ($r$). We trained several models with various number of residual blocks. More specifically, we varied $r$ from $1$ to $5$, and trained $5$ different smoothing networks. Our experiments show that MSE of the models with $r=3,4,5$ are very close, and the improvement from $r=1$ to $r=2$ is about $3.5\%$. Increasing $r$ from $2$ to $3$ only showed a modest $0.2\%$ improvement in the MSE.  

We conclude by referring to run time: We ran both our editors on an Intel Xeon CPU @ 3.5 GHz with 32 GB memory and 12 cores. We only measure timing of the pre-editing operation, as both methods use the same JPEG encoder. The smoothing CNN and STN run in $1.7$ sec and $1.2$ sec on a 1 mega pixel image, respectively. Since our editors are based on convolutional neural networks, these running times can be further improved by GPU inference. 


\section{Conclusion}
\label{sec:Conclusion}

One of the main bottlenecks of low bit-rate JPEG compression is loss of textures and details and presence of visual artifacts. In this work we have proposed an end-to-end trainable manipulation framework that edits images before compression in order to mitigate these problems. Our CNN-based trained editors optimize for better perceptual quality, lower JPEG distortions and color degradation. The proposed image-editors are trained offline, avoiding the need for per-image optimization, or a post-processing on the decoder (client) side. It is worth mentioning that JPEG is the dominating compression technology in digital cameras, cellphones, and webpages, with nearly $70\%$ of all websites on the internet using images with JPEG format~\cite{w3techs}. Thus, improvements in the JPEG standard may have significant industrial impact. Our future work will focus on extending this idea to other image compression standards, while seeking new ways to allow for more daring editing effects.


		

{\small
\bibliographystyle{ieee_fullname}
\bibliography{references}
}

\end{document}